\documentstyle[12pt]{article}
\newcommand{\sect}[1]{\section{#1}\setcounter{equation}{0}}

\gdef\journal#1, #2, #3, 1#4#5#6{
    {\sl #1~}{\bf #2} (1#4#5#6) #3}
\newcommand{\prl}{\journal Phys. Rev. Lett., }
\newcommand{\cmp}{\journal Comm. Math. Phys., }

\newcommand{\pr}{\journal Phys. Rev., }
\newcommand{\ap}{\journal Ann. Phys., }

\newcommand{\np}{\journal Nucl. Phys., }
\newcommand{\pl}{\journal Phys. Lett., }

\newcommand{\eq}{\begin{equation}}
\newcommand{\eqe}{\end{equation}}
\newcommand{\dis}{\begin{displaymath}}
\newcommand{\dise}{\end{displaymath}}
\newcommand{\eqa}{\begin{eqnarray}}
\newcommand{\eqae}{\end{eqnarray}}
\newcommand{\e}[1]{\label{eq:#1}}
\newcommand{\ee}[1]{(\ref{eq:#1})}
\newcommand{\RN}{Reissner-Nordstr\"om}

\begin{document}\bigskip
\hskip 3.7in\vbox{\baselineskip12pt
\hbox{UCSBTH-97-06}\hbox{gr-qc/9704072}}
\bigskip\bigskip\bigskip

\centerline{\large \bf Quantum States of Black Holes\footnote{To appear
in the proceedings of the symposium on Black Holes and Relativistic Stars,
in honor of S. Chandrasekhar, December 1996.}}

\bigskip\bigskip

\centerline{\bf Gary T. Horowitz}
\medskip
\centerline{Physics Department}
\centerline{University of California}
\centerline{Santa Barbara, CA. 93106}
\medskip
\centerline{email: gary@cosmic.physics.ucsb.edu}

\begin{abstract}
\baselineskip=16pt
I review the recent progress in providing a statistical foundation
for black hole thermodynamics. In the context of string theory, one
can now identify and count quantum states associated with black holes.
One can also compute the analog of Hawking radiation (in a certain
low energy regime) in a manifestly
unitary way. Both extremal and nonextremal black holes
are considered, including the Schwarzschild solution.
Some implications of conjectured non-perturbative
 string ``duality transformations''
for the description of black hole states are also discussed.
\end{abstract}
\newpage
\baselineskip=16pt

\sect{Introduction}

When I was a graduate student at the University of Chicago in the
late 1970's, Chandra often talked about the surprises he found in
his study of black holes \cite{chan}: the separation of the Dirac equation in
the Kerr metric, the equality of reflection and transmission
coefficients for different types of perturbations, etc. 
Given his well known fascination 
with black holes, I am sure Chandra would be interested in
the unexpected results about black holes described below,
which have been discovered in
the past year or two.

It is by now well known that black holes have
thermodynamic properties (for a recent
review see \cite{wald}). In particular, they have 
an entropy equal to one quarter of the area of their event horizon 
in Planck units.
This is an enormous entropy. One way to see this is to consider
thermal radiation (which, of course, has the highest entropy of ordinary matter)
and ask what entropy it would have when it forms a black hole. In
Planck units $G=c=\hbar=k=1$, a ball of radiation
at temperature $T$ and radius $R$ has mass $M\sim T^4 R^3$, and entropy
 $S\sim T^3 R^3$.
The radiation will form a black hole when $R\sim M$  which implies
$T \sim M^{-1/2}$ and hence $S\sim M^{3/2}$. In contrast, the entropy
of the resulting black hole is $S_{bh}\sim M^2$. So any black hole with
$M\gg 1$, i.e. mass much larger than the Planck mass, has an entropy
much larger than the entropy of the thermal radiation that formed it.
In terms of a more fundamental description, the entropy should be a measure
of the number of underlying 
quantum states. 
The problem, which has puzzled physicists for more than twenty years,
is to find a more fundamental theory which contains 
 the quantum states predicted by black hole thermodynamics.

An answer has recently been provided in the context of string theory.
This is a new physical theory which came into prominence in the mid
1980's. This theory has been hailed as a ``theory of everything" and
scorned as a ``typical end of the century phenomenon" \cite{anon}.
I think the first
view is much closer to the truth. String theory
is a promising candidate for a consistent quantum theory of gravity
and a unified theory of all forces and particles. It has not been proven
that it can achieve these goals, but there is increasing evidence
(especially over the past few years \cite{vafa}) that it will.

To understand the basic idea behind the recent explanation of
black hole entropy, one only needs
three facts about string theory \cite{gsw}. The first is that when one quantizes
a string in flat spacetime, there are an infinite tower of massive states.
For every integer $N$ there are states with $M^2 \sim N/l_s^2$ where $l_s$ is
a new length scale in the theory set by the string tension. These states
are highly degenerate, and one can show that
the number of string states at excitation level $N\gg 1$ is $e^{S_s}$ where
\eq
S_s \sim \sqrt N
\eqe
i.e. the string entropy is proportional to the mass in string units. One can
understand this in terms of a simple model of the string as a random walk
with step size $l_s$.\footnote{This model is surprisingly accurate: The
typical configuration of the string in a highly excited state is indeed a
random walk \cite{mitu}.} As a result of the string tension, the energy 
in the string after $n$ steps is
proportional to its length: $E\sim n/l_s$.
If one can move in $k$ possible directions at each step, the total number of
configurations is $k^n$, so the entropy for large $n$ is proportional to
$n$, i.e. proportional to the energy.

The second fact is that string interactions are governed by a string coupling
constant $g$ (which is determined by a scalar field called the dilaton).
Newton's constant $G$ is related to $g$ and the string length $l_s$
by $G \sim g^2 l_s^2$ in four spacetime dimensions. We will sometimes use
string units where $l_s=1$, and sometimes use Planck units where $G = l_p^2 =1$.
It is important to distinguish them, especially when $g$ changes.
 Since $g$ is in fact determined by a dynamical
field, one can imagine that it changes as a result of a physical process,
e.g. a wave of dilaton passing by. However, it will often be convenient
to assume the dilaton is constant and treat $g$ as just a parameter in 
the theory. In general, physical properties of a state can change
when $g$ is varied. But we will see that
in some cases, one can argue that certain 
properties remain unchanged.

The third fact is that the classical
spacetime metric
is well defined in string theory only when the curvature is less than
the string scale $1/l_s^2$. This follows from the fact that fundamentally,
the metric is
unified with all the other modes of the string. This is easily seen in
perturbation theory where the graviton is just one of the massless excitations
of the string. When the curvature is small compared to $1/l_s^2$,
one can integrate out the massive modes and obtain an effective low
energy 
equation of motion which  takes the form of Einstein's equation with an infinite
number of correction terms consisting of higher powers of the curvature
multiplied by powers of $l_s$.  When curvatures approach the string scale,
this low energy approximation breaks down.

In the next section we give a general argument which shows that 
string theory can explain the entropy of black holes. This argument
applies to essentially all black holes, and reproduces the correct dependence
on the mass and charges, but is not precise enough
to check the numerical coefficient. In section three we show that for
certain black holes, even the numerical coefficient in the entropy can
be computed and shown to agree. We also describe some recent calculations
showing that the spectrum of Hawking radiation from black holes can be 
reproduced by
string theory. In section four, we consider the effect of conjectured
nonperturbative ``duality" symmetries  on the description of black hole states. 
Section five contains a discussion of some open issues.

\sect{Correspondence between black holes and strings}

\subsection{Schwarzschild Black Holes}

A few years ago, Susskind \cite{suss1}
suggested that there should be a one to one
correspondence between strings and black holes. The idea was the following:
Consider a  highly excited string state at level $N$ and
zero string coupling. As we mentioned above, 
a typical
configuration of such a string is a random walk with a length  proportional
to its energy $L\sim N^{1/2} l_s$ and hence a radius $R\sim N^{1/4} l_s$. 
 Now imagine
increasing the string coupling $g$ and recall that $G \sim g^2 l_s^2$.
Two effects take place. First,
the gravitational attraction of one part of the string on the other
causes the string size to  decrease. 
Second, since $G$ increases, the gravitational field produced by the string
becomes stronger and the effective
Schwarzschild radius  $GM$ increases in string units. Clearly, for a 
sufficiently large  value of the
coupling, the string forms a black hole. 

Conversely, suppose  one starts with a black hole and decreases the
string coupling. Then the Schwarzschild radius
shrinks in string units and 
eventually becomes smaller than the string scale. At this point
the metric is approximately flat except for a small region where it is
no longer well defined. Susskind suggested that the 
black hole becomes an excited string state. Further evidence was presented
in \cite{hkrs}.

When I first heard this, I didn't believe it. The first half of the argument
sounded plausible enough, but the second half seemed to contradict the
well known fact that the string entropy is
proportional to the mass while the black hole entropy is proportional to the
mass squared. It turns out that there is a simple resolution of
this apparent contradiction \cite{hopo}.
Consider the familiar Schwarzschild black hole
\eq \e{schw}
ds^2 =-\left( 1 - {r_0\over r}\right) dt^2 + \left( 1 - {r_0\over r}\right)^{-1}
dr^2 + r^2 d\Omega\ .
\eqe
The mass of the black hole is $M_{bh} = r_0/2G$. We want to equate this
with the mass of a string state at excitation level $N$ which is $M_s^2 \sim
N/l_s^2$ at zero string coupling. Now imagine increasing the string coupling,
keeping the state fixed.
Clearly, $M_s$ is constant in string units
where $l_s$ is held fixed and Newton's constant is changing\footnote{There
is  a small correction to $M_s$ due to gravitational binding energy, but
this is negligible compared to the effect discussed here.}. The analog of 
keeping the state fixed for the black hole is to keep the entropy 
fixed\footnote{For a system with a rapidly growing density of states,
a narrow band of
states can be labeled by its entropy. The argument below can be reinterpreted
as saying that if you insist on keeping the entropy constant through the
transition from strings to black holes, the mass changes by at most a
factor of order unity.}.
Thus $M_{bh}$
is constant in Planck units where $G$ is fixed and the string length changes.
(The black hole does not know about $g$ or $l_s$ separately, but only 
the combination $G\sim g^2 l_s^2$.) In
other words, the ratio $M_s/M_{bh}$ depends on $g$ and the masses cannot
be equal for all values of the coupling. If we want to equate the masses,
we have to decide at what value of the coupling they should be equal. Clearly
the natural choice is to let $g$ be the value at which the string forms
a black hole or vice versa.
 If we start with a black hole and decrease the coupling
the black hole description
will remain valid until the horizon is of order the string
scale.
Setting the masses equal when $r_0 \sim l_s$ yields
\eq
M_{bh}^2 \sim {l_s^2 \over G^2} \sim {N\over \l_s^2} \e{match}
\eqe
The black hole entropy is then
\eq\e{swent}
S_{bh} \sim {r_0^2 \over G}\sim {l_s^2 \over G} \sim \sqrt N
\eqe
So the Bekenstein-Hawking entropy is comparable to the string
entropy! 
One cannot compare the numerical coefficient in the entropy formulas
this way, since that would depend on exactly when the transition occurred.
But this clearly shows that  strings
have enough states to reproduce the entropy of black holes.
It follows from \ee{swent} that
the transition from a string state to a black hole occurs when the
string coupling is still rather small: $g\sim 1/N^{1/4}
\ll 1$ for large $N$.

Let me emphasize that the fact that $r_0 \sim l_s$ does not mean that
the black hole must be small. In fact, since the entropy is $S_{bh} \sim
\sqrt N$, the Schwarzschild radius in Planck units is $r_0 \sim N^{1/4}$.
It is probably more intuitive to describe
the transition between  string states and black holes 
in Planck units where $G$ is fixed. Suppose we start with a solar
mass black hole. As we decrease $g$, the string length
$l_s$ increases, i.e., the string tension decreases.
Eventually, it is of order a kilometer and comparable to the horizon size.
At this point, the black hole metric is no longer well defined in
string theory, and the system is  better described as a highly
excited string state.

Starting at small coupling, the transition from a
string to a black hole is analogous to 
a neutron star which accretes matter. 
In Planck units, the string tension increases like $g^2$. Thus for fixed
excitation $N$, the mass of a string state {\it  increases} as $g$ increases
until it forms a black hole.
For both the neutron star and the string, there is not a sudden change in mass
when the black hole forms.
However there is one important difference: Neutron stars have  much less entropy
than the resulting black hole, while strings do not.

The above counting of string states does not include the center of mass
degree of
freedom. This is appropriate since the entropy of a black hole does not
include a contribution from its location in spacetime. It also does not include
the possibility that a black hole is described by several strings at
weak coupling.  Since the string entropy is proportional to its
energy, two strings with energy $E/2$ have
the same entropy as one string with energy $E$. However, it turns out
that the ratio of the total entropy of all
multistring states to the entropy of a single string goes to one for large
$E$ \cite{mitu}. So the entropy can be approximated by a single string.

We have seen that as one increases $g$, a perturbative string state
evolves into a black hole.
In addition to the string-black hole transition,
there seems to be an implicit quantum-classical transition as well.
This can be made explicit as follows. Recall that for a given excitation $N$, 
the transition between the black hole and string regimes
occurs when $g N^{1/4} \sim 1$. For $g N^{1/4} > 1$ the
Schwarzschild radius is larger than the string scale and the black hole
description is valid. If $g N^{1/4}<1$, the
effective Schwarzschild radius is less than the string scale and
perturbative string description is valid. {\it In either case,} the
classical limit is $g\rightarrow 0,\ N\rightarrow \infty$ with
$g N^{1/4}$ fixed. In this limit, the Planck length $l_p \sim g l_s$
goes to zero, and the entropy $S \sim \sqrt N$ diverges, which agrees
with the fact that a purely classical black hole indeed has infinite
entropy. Quantum corrections to the black 
hole can be computed in terms of a string loop expansion. 
The classical limit on the
perturbative string theory side does not yield a single classical string.
Since string theory is a ``second quantized" theory of strings, the states of
the first 
quantized string describe the classical fields of the final theory. 
So the $g\to0$ limit is essentially a classical field theory with an
infinite number of fields. From this viewpoint,
it is interesting to note that the
black hole entropy is related to the {\it number of fields} in the theory
with a given mass. (Since many of these fields have large spin, one is
actually counting the number of components of these massive fields.)
In the second quantized theory, the first quantized string
states simply arise as states in the ``one string sector" of the
usual perturbative 
Fock space. 
In the following, we will often keep $N$ fixed and vary $g$. Thus, the
black hole will be referred to as the ``strong coupling regime".

The agreement between black hole entropy and the number of string
states  extends to Schwarzschild black holes in higher dimensions.
The easiest way to see this is to note that the entropy of a
Schwarzschild black hole in any dimension can be expressed 
\eq
S_{bh} \sim r_0 M_{bh}
\eqe
Clearly, when $r_0 \sim l_s$, the entropy
is proportional to the mass in string units, exactly like a free string.

\subsection{A Charged Black Hole}

The agreement between the black hole entropy and the counting of
string states also extends to charged black holes. The simplest
case to consider is a Kaluza-Klein black hole where the charge comes from 
momentum in an internal direction \cite{kk}. 
Given  a five dimensional metric which is
independent of $x_5$, define a four dimensional metric $g_{\mu\nu}$,
gauge field $A_\mu$ and scalar $\chi$ by
\eq\e{kkr}
ds^2 = e^{-4\chi/\sqrt 3} ( dx_5 + 2 A_\mu dx^\mu)^2
+ e^{2\chi/\sqrt 3} g_{\mu\nu} dx^\mu dx^\nu 
\eqe
Then the five dimensional Einstein action is equivalent to
\eq
S= \frac{1}{16\pi G} \int d^4x \sqrt{-g}[R - 2(\nabla \chi)^2 - 
e^{-2\sqrt 3 \chi}F_{\mu\nu} F^{\mu\nu}]
\eqe
Charged black hole solutions can
be obtained by starting with the product of the four dimensional 
Schwarzschild metric \ee{schw} and a line, boosting along the line,  
periodically identifying $x_5$ and reducing to four dimensions using 
\ee{kkr}. The result is
\eq\e{kkbh}
ds^2 = -\Delta^{-1/2}\left( 1-{r_0\over r}\right) dt^2 +
 \Delta^{1/2}\left [\left( 1-{r_0\over r}\right)^{-1} dr^2  + r^2 d\Omega
 \right ]
  \eqe

  $$A_t =-\frac{r_0\sinh 2\gamma}{4r\Delta}, \qquad \qquad 
e^{-4\chi/\sqrt{3}} = \Delta$$
  where
  $$ \Delta = 1+{r_0 \sinh^2 \gamma\over r} $$
The ADM mass and electric charge are
\eq
M_{bh}= {r_0\over 8G} (3+\cosh 2\gamma)
\eqe
\eq
Q= {r_0\over 4G} \sinh 2\gamma
\eqe
The black hole entropy is
\eq\e{kkent}
S_{bh} = {\pi r_0^2\over G} \cosh \gamma
\eqe
We want to show that this entropy is reproduced by  counting  string
states. The electric charge on the black hole is twice the total momentum 
in the internal direction. A string in five dimensions with momentum $P$ 
in the fifth direction  
has entropy $S_s \sim \sqrt N$ where now 
\eq\e{kkstates}
M_s^2 - P^2 \sim {N\over l_s^2}
\eqe
and $M_s$ is the four dimensional mass of the string. We want to match
the black hole solution to the string solution so we set $P=Q/2$. We also
set $M_{bh}=M_s$ when the curvature is
of order the string scale. It is the curvature of the original five dimensional
solution which is important here so the matching occurs when $r_0 \sim l_s$.
The result is that

\eq\e{ndef}
{N\over l_s^2} \sim  M_{bh}^2 - {Q^2\over 4} \sim {l_s^2 \over G^2}
[5+3\cosh 2\gamma] 
\eqe
Comparing with \ee{kkent} we see that up to a ($\gamma$ dependent)
factor of order unity,
\eq
S_{bh} \sim \sqrt N \sim S_s
\eqe
Thus the two entropies agree for all values of $\gamma$, i.e. for all
values of $Q/M$.

\subsection{Correspondence Principle}

For the Kaluza-Klein black hole,
we assumed there was an internal direction in spacetime. String theory
actually predicts many internal directions. (In the above
examples, the extra dimensions were taken to be a trivial torus.)
These give rise to a large number of different types of charges 
in the effective four dimensional theory. In addition to internal momentum,
one has charges associated with strings winding around each compact direction.
There are also higher dimensional extended objects called D-branes
(which will be discussed in the next section) which look like localized
charged particles when wrapped around the internal space. Black holes
can carry any of these charges. In almost all cases, the size of the
black hole in string units becomes smaller when $g$ is decreased.\footnote{The
one exception is black holes carrying Kaluza-Klein monopole charge, or
another magnetic charge dual to string winding  number.  Approaches to
understanding the entropy of black holes with these charges have been discussed
in \cite{lawi,cvts,dvv1}.} 
The entropy of all of these black holes can be understood
by matching onto a weak coupling description when the black hole is of
order the string scale. However, there is one additional fact about
string theory which must be taken into account.
For many charged black holes, the dilaton field $\phi$ is not
constant. In this case, strings couple to a metric $g_{\mu\nu}^S$ which
is  conformally related to the metric with the standard Einstein action 
$g_{\mu\nu}^E$. In $D$ spacetime dimensions, one has 
$g_{\mu\nu}^S = e^{4\phi/D-2} g_{\mu\nu}^E$. 
Both of these metrics play a role in our discussion.
The spacetime metric ceases to be well defined when the curvature of the string
metric is of order the string scale, while the black hole entropy
is, of course, related to the area of the event horizon in the standard
Einstein metric. 

The general relation between black holes and strings
can now be stated in terms of the following {\bf correspondence principle}:

$(i)$ When the curvature at the horizon of a black hole (in the string
metric) becomes greater than
the string scale,  the typical black hole state becomes a typical state
of strings and D-branes with the same charges and angular momentum.

$(ii)$ The mass changes by at most a factor of order unity during the
transition.

\noindent An appropriate measure of the curvature near the horizon are the
curvature invariants, since these control the higher order string corrections
to the field equations. 
(The physical tidal forces felt by an infalling observer can be
much larger than the curvature invariants would suggest \cite{horo}.) 
It has been shown \cite{hopo} that for a large class of charged black holes, 
this
correspondence principle correctly reproduces the Bekenstein-Hawking entropy 
up to factors of order unity. The two examples discussed earlier are easily
seen to be special cases of this principle. 

As a final example, consider
the following metric describing a five dimensional  black string 
\cite{host1}
\eq\e{bkst}
ds^2 = F\left[ -\left( 1-{r_0\over r}\right) d t^2 + d z^2\right]
+ \left( 1-{r_0\over r}\right)^{-1} dr^2 + r^2 d\Omega
\eqe
where
\eq
F^{-1}= 1+ {r_0 \sinh^2 \gamma_1\over r}\ .
\eqe
This is not the Einstein metric, but the string metric $g^S_{\mu\nu}$.
The dilaton is $e^{2\phi} = F$.
The extremal limit, $r_0 \rightarrow 0,   \ \gamma_1 \rightarrow \infty$ with
$r_0 \sinh^2 \gamma_1$ fixed, describes the strong coupling geometry of
an unexcited string wrapped around
an internal
direction. 
When this solution was discovered, it was suggested that
the nonextremal black string should describe the strong coupling limit
of the excited states of the wound string. We now see that this is indeed
the case. In fact, \ee{bkst}
reduces to the same four dimensional black hole 
that we discussed earlier \ee{kkbh},
and the counting of states for a string with
nonzero winding number is
again given by \ee{kkstates} with $P$ replaced by the energy in the
winding mode. So the number of string states 
with nonzero winding number agrees with the black string
entropy.

\sect{Precise agreements between black holes and strings}

\subsection{Supersymmetric Black Holes}

Using the correspondence principle to understand black hole entropy
has the virtue that it can be applied to 
essentially any black hole, but it is not yet able to compare the
precise coefficients in the entropy formulas. For a special class of
black holes, as one decreases the string coupling,
one  has more control over the transition to a perturbative string
state and even the coefficients in the entropy formulas
can be compared. These more precise calculations use the fact that
string theory is supersymmetric. Its low energy limit is a supergravity
theory that admits black hole solutions  which are invariant under
some supersymmetry transformations. These 
supersymmetric solutions are extremally charged black holes with mass
and charge related by $M=cQ$ for some constant $c$.
In the limit of weak coupling, the mass and charge of
all perturbative string states 
satisfy inequalities of the form\footnote{This is only true for certain
charges in extended supersymmetry which appear in the supersymmetry algebra.}
$M\ge cQ$.
States with $M= cQ$ are called BPS states, and have the special property
that their mass does not receive any quantum corrections.

One can thus start with a supersymmetric black hole, and decrease
the string coupling. One then counts the number of BPS states
at weak coupling with the same charge as the black hole. Notice that
in this case, the issue of when to match the mass of the black hole
to the perturbative string state never arises: In both regimes
the mass is completely fixed by the charge. The remarkable
result is that the number of BPS states at weak coupling
turns out to be precisely the exponential of the Bekenstein-Hawking
entropy of the black hole at strong coupling \cite{stva}. 
(For a comprehensive review, see \cite{malthe}.)

This sounds so easy, one might wonder why this calculation
wasn't done years ago. There are two main reasons. The first is that
most supersymmetric black holes  are not really black holes at all:
They have zero horizon area. The problem is
that supergravity theories have scalars which couple to the gauge fields.
Nonextremal charged black hole solutions exist with familiar properties, but
as one approaches the extremal limit, the scalars become large at the
horizon causing it to shrink to zero size.
To obtain a supersymmetric black hole with nonzero horizon area, one
needs to include several charges to stabilize the scalars. This 
results in a second problem, since some of these
charges are not carried by fundamental strings.  Instead, they
are carried by nonperturbative solitons. Thus, rather than simply counting
states of a string at weak coupling, one must quantize the solitons
and count bound states of the solitons and strings, which is much
more difficult. These problems were recently solved
when
Polchinski discovered a new representation for these solitons called
``D-branes" \cite{pol}. 

A D-brane has a mass proportional to $1/g$, so at weak coupling they
are very massive (and hence nonperturbative). But since Newton's constant
$G\sim g^2$, the gravitational field produced by the D-brane, which
is proportional to  $GM$ goes to 
zero as $g\rightarrow 0$. Thus there exists a flat space description of
these nonperturbative states. This is obtained by adding
to one's theory of closed
strings, a set of open strings where the endpoints of the open strings
are constrained to live on a  particular surface. These surfaces can have
any dimension, and can be viewed as generalizations of membranes.
Since the endpoints
of the open string satisfy Dirichlet boundary conditions in the directions
normal to the surface, the surface is called a D-brane. 

The excited states of a D-brane are described by quantizing the
open strings. At low energies, only the massless modes contribute.  The massless
states of an open string include some scalars which describe the fluctuations
of the brane in the surrounding spacetime, and a $U(1)$ gauge field on the
brane.
When two D-branes
come together this is enhanced to a $U(2)$ gauge field. The extra
massless states needed to change $U(1)^2$ into $U(2)$
arise from open strings that are stretched between
the two D-branes. These states have a mass proportional to the
separation of the branes. When this separation goes to zero, they become
massless and combine
with the $U(1)$ gauge fields on each brane to produce a $U(2)$ gauge field.
Similarly, when $m$ D-branes come together one obtains a $U(m)$ gauge
theory.  
In terms of the low energy gauge theory on the brane, the fact that
$U(m)$ reduces to $U(1)^m$ when the branes are slightly separated 
is described by the usual Higgs mechanism. 

There is a fascinating story which is emerging from the study
of D-branes at very short distances. I do not have time to develop it
in detail (and it is not required to understand the recent progress
in black holes) but I cannot resist mentioning it. It appears that
D-branes are able to probe distances much shorter than the string
scale \cite{dkps}. Since the graviton is a mode of a closed
string, the usual metric
description cannot apply at these short distances. Consider
$p$-dimensional D-branes.  When $m$ of them come together, their
low-energy dynamics is governed by the reduction of a ten dimensional
$U(m)$ Yang-Mills theory to $p$ dimensions.
Thus, in addition to the $U(m)$ gauge field on the brane there
are matter fields $X_i, \ i=p+1,\cdots, 9$ which are $m\times m$ hermitian
matrices with a potential $V\sim [X_i, X_j]^2$. For the ground states,
$[X_i, X_j]=0$, and one can simultaneously diagonalize these matrices.  When
$X_i\not= 0$ the symmetry is broken to $U(1)^m$ and the diagonal entries
can be interpreted as the positions of the (now separated) branes.
 Therefore,
there is a one-to-one relation between the position of
the D-brane in spacetime and the moduli space of
ground states of the gauge theory on
the D-brane. For slowly moving D-branes there is a natural metric on
this moduli space which controls the physics. In some cases, this
metric turns out to be
identical to the metric on spacetime measured at much larger
distances. Perhaps the most intriguing observation is that
the variables
in the gauge theory which  correspond to position in spacetime 
commute only for the ground states, but in general are noncommuting! This
suggests that at very short distances, spacetime may be described
by a form of noncommutative geometry \cite{con}.

Returning to our discussion of black holes,
a single D-brane with $p$ spatial dimensions 
carries a charge with respect to a $p+2$ form field strength $F$.
In other words, the charge is defined by integrating ${}^*F$ over a sphere
encircling the $p+1$ dimensional world volume of the brane.
In fact, a D-brane has the minimum possible mass for this type of charge
and is (in a well-defined sense) a BPS state.
Now suppose one compactifies spacetime on a $p$ dimensional internal space, and
wraps the D-brane around this space. From the standpoint of the reduced
theory, the D-brane appears to be a point-like object carrying a charge
associated with a usual two-form Maxwell field. Since a single D-brane
carries only one type of charge,  its strong coupling limit does not
have nonzero horizon area. However, bound states of several different
types of D-branes
yield black holes with regular horizons.

It turns out that supersymmetric
black holes with nonzero horizon area can only exist in four and five
dimensions. In five dimensions one needs three different types of charges
while in four dimensions, one needs four.\footnote{More precisely, this
is true in $N=4$ or $N=8$ supergravity. In $N=2$ supergravity, there are
supersymmetric black holes with one charge and nonzero horizon area \cite{gihu}.
However, this theory does not arise when compactifying string theory on a
torus. One needs  more complicated internal spaces \cite{n2bh}.}
The first precise calculation of black hole entropy was performed by
Strominger and Vafa \cite{stva} for an extreme
nonrotating five dimensional black hole. 
This was quickly generalized to include rotation \cite{bmpv} and 
four dimensional extremal black holes \cite{mast1,jkm} (as well as
small deviations from extremality which will be discussed shortly). 
Even the entropy of solutions which depend on arbitrary functions
has been reproduced by counting states of D-branes\footnote{In this case there
is
a small puzzle since although the horizon area is well defined,
the curvature diverges there \cite{hoya}.} \cite{homa}.

As perhaps the most interesting example of the above results, we consider
the familiar extreme \RN\  solution:
\eq \e{rn}
ds^2 =-\left( 1 - {GM\over r}\right)^2 dt^2 +
\left( 1 - {GM\over r}\right)^{-2}
dr^2 + r^2 d\Omega\ .
\eqe
As explained above, the counting of states for this black hole
is rather complicated since we need to consider four different charges. 
In other words, one views \ee{rn} as a composite of four different
fundamental objects. There is a more general solution (given below) where these
charges are all independent parameters, which reduces to \ee{rn} in a
certain limit.
Many of these charges are represented in weak coupling by D-branes wrapped
around internal directions. There are, in fact,
several different possible choices for the charges
which all include \ee{rn} as a special case.

One way to obtain the \RN\ solution is the following \cite{mast1}.
One starts at weak coupling
with  ten dimensional flat spacetime and compactifies six dimensions
on a torus, which is convenient to think of as the product
of a four torus with volume $(2\pi)^4 V$ and two circles with
radii $R$ and $\tilde R$. One then takes $Q_6$ D-sixbranes wrapped around
the six torus. One adds $Q_2$ D-twobranes wrapped around the two circles.
One adds $Q_5$ five branes wrapped around the four torus and $R$.  All
of these branes lie over the same point in the three noncompact spatial
directions, so this configuration
describes a localized object in four spacetime dimensions.
Notice
that the intersection of these branes is the circle $R$. When $R$
is large the low energy excitations of these branes are described by open
strings moving along this circle. It turns out that these strings have
$4Q_2 Q_5 Q_6$
massless bosonic degrees of freedom and an equal number of fermionic
degrees of freedom. (This is obtained by analyzing the induced gauge theory
on the branes.) Momentum along the circle is quantized in units of $1/R$.
The number of states with right-moving momentum  $n/R$ (and no left-moving
momentum) is $e^S$ where $S$ is given by
the general formula for a one dimensional
gas $S= 2\pi \sqrt{cn/6}$. The constant $c$  receives a contribution of
one for every bosonic field and one half
for every fermionic field. 
In our case, we have $c= 6Q_2 Q_5 Q_6$
and hence
\eq\e{stent}
S=  2\pi \sqrt{Q_2 Q_5 Q_6 n}
\eqe
I should emphasize that this calculation is done in flat spacetime. There
is no event horizon present. One is simply counting states of strings on
D-branes.
Notice that the entropy depends only on the integer charges, and
not on the continuous parameters $V, R, \tilde R$.

The strong coupling description of this system is found by solving the ten
dimensional supergravity equations with these charges. After reducing
along the six torus, the four dimensional (Einstein) metric becomes\footnote{We
follow the discussion in \cite{hlm}.}
\cite{cvyo} 
$$ds^2 = - f^{-1/2}(r)\left(1 - {r_0  \over r}\right) dt^2 +
 {f^{1/2}(r) }\left[
 \left( 1 -{r_0 \over r} \right)^{-1}  dr^2
 +  r^2 d\Omega \right]~,$$
where
\eq\e{grn}
 f(r) =
 \left(1 +{ r_0 \sinh^2
 \alpha_2  \over r}\right)\left(1 + { r_0  \sinh^2 \alpha_5 \over r}\right)
 \left(1+ { r_0  \sinh^2 \alpha_6 \over r}\right)\left(1 + { r_0
 \sinh^2 \alpha_p \over r }\right)~.
 \eqe
This metric is parameterized by the
five  independent quantities $\alpha_2$, $\alpha_5$,
$\alpha_6$, $\alpha_p$ and $r_0$. They are related to the integer
charges by
$$Q_2  = {  r_0 V \over g }
\sinh 2\alpha_2 ~, $$
$$Q_5 = { r_0 \tilde R   }
\sinh 2\alpha_5 ~, $$
$$ Q_6 = {  r_0 \over  g}
\sinh 2\alpha_6 ~, $$
\eq\e{charges}
n = {  r_0 V R^2  \tilde R \over g^2 }
\sinh 2\alpha_p ~, 
\eqe
where we have set $l_s=1$.
The event horizon lies at $r=r_0$. The special case $\alpha_2 = \alpha_5
=\alpha_6=\alpha_p$ corresponds to the Reissner-Nordstr\"om metric.
Notice that if we set all charges except the momentum $n$ to zero,
the metric \ee{grn} reduces to the Kaluza-Klein solution \ee{kkbh} as
it should. Furthermore, since all charges enter \ee{grn} symmetrically,
the four dimensional metric generated by any one of these charges is the
Kaluza-Klein black hole. The precise relation between
the four-dimensional Newton constant and the string coupling is
$G =  g^2/(8 V R \tilde R)$ in string units ($l_s =1$).
The ADM mass of the solution is
\eq\e{admmass}
M= { r_0  V R \tilde R \over g^2}
 (\cosh 2\alpha_2+
 \cosh 2\alpha_5 + \cosh 2 \alpha_6 + \cosh 2 \alpha_p )
\eqe
and the Bekenstein-Hawking entropy is 
\eq\e{bhentropy}
S_{bh} ={A\over 4G} = {8 \pi r_0^2 V R \tilde R \over g^2} \cosh \alpha_2
\cosh \alpha_5 \cosh \alpha_6 \cosh \alpha_p ~.
\eqe
The extremal limit corresponds to $r_0 \rightarrow 0, \
\alpha_i \rightarrow \pm \infty$ with $Q_i$ fixed. In this limit,
the entropy becomes
\eq
S_{bh} = 2\pi \sqrt{Q_2 Q_5 Q_6 n}
\eqe
in precise agreement with the string result \ee{stent}! The \RN\ solution
is clearly just a special case of this.

Even though we have not needed the correspondence principle to 
reproduce the black hole entropy, one can still use it to
estimate when the black hole description breaks down, and must be
replaced by the D-brane description. In string units, the area of the
event horizon is approximately $g^2 \sqrt{Q_2 Q_5 Q_6 n}$. The curvature
at the horizon will be of order the string scale when this is of order one.
If we assume all the charges are comparable, the transition occurs when
$gQ\sim 1$.

In the extremal limit, the ADM mass becomes
\eq\e{exmass}
M = {R \tilde R\over g} Q_2 + {R V\over g^2} Q_5 + {R\tilde R V\over g} Q_6
+ {n\over R}
\eqe
which is easily seen to be just the sum of the energy of each of the four
constituents making up the black hole.\footnote{The reason that the energy
of the fivebranes goes like $1/g^2$ rather than $1/g$ is because 
it is not
a D-fivebrane but rather a solitonic fivebrane \cite{mast1}. Fortunately,
the counting of states can still be carried out in this case.
For an alternative description
of the four dimensional black hole without solitonic fivebranes, see
\cite{bala}.}
In the case of the \RN\ solution
when all the boost parameters are set equal, one can easily verify that
each of these constituents contributes equally to the total mass.
Notice that if one fixes the charges and size of the internal torus, the
mass (in string units) changes with the string coupling. This is consistent
with the BPS bound.

\subsection{Near Extremal Black Holes}

Since supersymmetry seemed to play an important role in the above 
discussion, one might have thought that the entropy could be calculated
precisely only in this case. This turns out to be incorrect. In fact,
soon after the first precise calculation of black hole entropy, it was
shown that 
the entropy of certain slightly nonextremal black holes
can also be calculated exactly \cite{cama,host2}.  At present, this is
well understood only in the
regime where one of the constituents is much lighter than the rest. This 
case is clearly the simplest to consider since
the maximum entropy is obtained by adding energy to the 
lightest degrees of freedom. For the case of \RN, where all the branes
contribute equally to the mass, if one adds a small amount of energy,
the excitation of
all the branes will contribute equally to the entropy, and the counting
is much more difficult.	It is clear from \ee{exmass} that when $R$ is large,
the momentum modes are
much lighter than the branes. So if one adds a small amount of excess
energy, it simply goes into exciting left and right moving modes. Since
the left and right moving modes are largely
noninteracting at weak coupling, the entropy is additive and we obtain
\eq
S = 2\pi \sqrt{Q_2 Q_5 Q_6 }(\sqrt{n_L} + \sqrt{n_R})
\eqe
where the left and right moving momenta are $n_L/R$ and $n_R/R$
respectively. This agrees precisely
 with the Bekenstein-Hawking  entropy computed
 from the near extremal hole solution \ee{bhentropy}. 
For other choices of the parameters $V, R, 
\tilde R$,
one of the branes may be much lighter than the other constituents. 
In this case, the
entropy can again
be understood by a suitable counting argument.
But when several branes are light (or all contribute equally, as in \RN),
one does not yet have a precise counting of the entropy of
the near extremal solution.

At weak coupling the interactions between the left and right moving
modes  are small but nonzero. Occasionally, a left-moving mode
can combine with a right-moving mode to form a closed string which
can leave the brane. This represents the decay of an excited configuration
of D-branes and is the weak coupling analog of Hawking radiation.
Given the remarkable agreement between the entropy of the black hole
and the counting of states of the D-branes, the next step is clearly 
to ask how the radiation emitted by the D-brane compares to Hawking
radiation.  
Since the entropy as a function of energy
agrees in the two cases, it is not surprising that the radiation is
approximately thermal with the same temperature in both cases. What
is surprising is that the overall rate of radiation agrees \cite{dama}.
What is even more remarkable is that the deviations from the
black body spectrum also agree \cite{mast2}. 
On the black hole side, these deviations
arise since the radiation has to propagate through the curved spacetime
outside the black hole. This produces potential barriers which give
rise to frequency-dependent greybody factors. On the D-brane side
there are deviations since the modes come from separate 
left and right moving
sectors on the  D-branes. The calculations of these deviations
could not look more
different.
On the black hole side, one solves a wave equation in a black hole 
background. The solutions involve hypergeometric functions. On the
D-brane side, one does a calculation in D-brane perturbation theory.
Remarkably, the answers agree.

To be a little more precise, the calculations were first done
for the (five dimensional)
near-extremal black hole with $R\gg 1$. Since the black hole is near
extremality, the temperature is very low and hence the wavelength of the
radiation is much larger than the size of the black hole.  One considers
radiation by a  minimally coupled scalar field.  On the $D$-brane side,
one starts with a thermal distribution of left and right moving modes
with temperature $T_L$ and $T_R$.  The decay rate for  
left and right moving
excitations, each  with energy $\frac{\omega}{2}$, to produce an
outgoing S-wave mode with energy $\omega$ is 
\eq\e{dbrt}
\Gamma ={g^2_{eff}\ \omega\over (e^{\omega/2T_L} -1)(e^{\omega/2T_R} -1)} 
{d^4 k\over
(2\pi)^4}
\eqe
The factors in the denominator are the usual thermal factors for the 
left and right moving modes and $g_{eff}$ is a frequency independent 
effective coupling constant. To compare with the black hole, one computes
the average left and right moving energy,  which determines $n_L$ and
$n_R$. This fixes the total energy and momentum of the black hole. 
The Hawking temperature
turns out to be related to $T_L$ and $T_R$ by
\eq
{1\over T_R} + {1\over T_L} = {2\over T_H}
\eqe
The black hole decay rate is given by \cite{haw}
\eq\e{bhrt}
\Gamma = \frac{\sigma_{abs}(\omega)}{(e^{\omega/T_H}-1)}
\ \frac{d^4k}{(2\pi)^4}
\eqe
where $\sigma_{abs}(\omega)$ is the greybody factor, which equals
the classical absorption cross section. One calculates
$\sigma_{abs}(\omega)$ by studying solutions to the wave equation in the
black hole background.  For the Schwarzschild and Kerr metrics, this was
extensively studied more than twenty years ago \cite{stch}.
It has recently been shown that for any black hole, the limit
of $\sigma_{abs}(\omega)$ as $\omega\rightarrow 0$ is the area of the event
horizon \cite{dgm}.   After a lengthy calculation in the metric analogous to
\ee{grn} describing a five dimensional black hole with three charges,
one finds \cite{mast2} that for 
$\omega \leq T_H$,
\eq
\sigma_{abs}(\omega)  
= {g_{eff}^2 \ \omega\ (e^{\omega/T_H}-1) \over (e^{\omega/2T_L}-1)
(e^{\omega/2T_R}-1)}
\eqe
so  the two rates \ee{dbrt} and \ee{bhrt}
agree!

It is worth emphasizing that it is not just a few parameters which agree.
Since the calculation is valid for $\omega \leq T_H$, the entire
functional form is significant.  It is as if the black hole 
knows that its states are described by an effective $1+1$ dimensional
field theory with left and right moving modes. It
appears that the black hole also knows that some of these modes are
fermionic. 
In the weak-coupling description, an outgoing mode with angular momentum
$\ell=1$ arises from a left and right moving fermion on the
D-brane. Remarkably, when the greybody factor is computed for this
case, one again finds that it factors into  left and right moving
thermal factors, but now they take the form $(e^{\omega/2T}+1)$ appropriate for
fermions \cite{mast3}! (The overall numerical coefficient has not yet
been checked
for this case).  More generally, for $\ell$ odd, one obtains the
fermionic factors, while for $\ell$ even, one obtains the bosonic factors
as expected from the D-brane description.

It
is not yet clear why a calculation of decay rates at weak coupling can be
extrapolated without modification into the strong coupling regime. One
possible explanation was given by Maldacena \cite{mald} who argued that at low
energy, the relevant interactions did not receive quantum corrections due
to a supersymmetric non-renormalization theorem (see also \cite{das}).

Attempts to extend this calculation have met with mixed success.
Agreement was found for four, as well as five, dimensional black holes 
\cite{gukl}, charged scalars \cite{mast2,gukl},  and
certain non-minimally coupled scalars \cite{cgkt,krt} but disagreement was
found for higher energies \cite{deal,dkt}, other near-extremal
regimes, e.g. $R\sim 1$ \cite{hata,dkt}, or other non-minimally-coupled
scalars \cite{krkl}.
However, in  most of the cases where disagreement was found,
even the weak-coupling D-brane
calculations are not yet completely understood.

These results have immediate implications for the well known
black hole information puzzle. Hawking has argued that the radiation
emitted from a black hole is independent of what falls in. Thus if an
extreme black hole absorbs a small amount of matter and then radiates back
to extremality, information will be lost, and unitarity will be
violated.
String theory provides a manifestly unitary description of a
system of strings and D-branes with the same entropy and rate of radiation.
If one throws a small amount of energy toward a system of D-branes, the
branes become excited and then decay. The final system is in a pure state,
with correlations between state of the D-branes and the radiation. If
one traces over the D-brane states, the radiation is approximately
thermal.
One might imagine that the same thing happens for black holes. But this
does not avoid the difficulties of how information gets out from inside
the black hole. 
Suppose that one repeats this experiment many times. When the entropy in
the radiation becomes greater than the entropy in the D-branes, it can
no longer be the case that the radiation is thermal  after tracing out
the D-brane states. There will be  correlations between the radiation
emitted at early times and the radiation emitted at late times \cite{page2}.
Since this system of strings and D-branes becomes a black hole when
one increases the string coupling, 
it is very tempting to conclude that even in the black hole regime, there
will be correlations between the radiation at early and late times 
and the evaporation process will be unitary. 
In fact, the argument based on supersymmetric
nonrenormalization theorems \cite{mald} supports this view
for very low energy quanta.

However, Hawking has stressed that the causal structure of the black hole
is very different  from the flat space description, since there is no
analog of the event horizon. It is logically possible that the evaporation
process is unitary at weak coupling and fails to be unitary at strong
coupling. Before one can conclude that quantum mechanics is not
violated in  black hole evaporation, one needs at least a convincing
explanation of where Hawking's original arguments break down.

\sect{Duality}

In the previous two sections,
we have considered the transition from a weakly coupled
state of strings and D-branes to a black hole. We have seen that
this occurs when $g N^{1/4}\sim 1$ or $gQ \sim 1$ which implies
that the fundamental
string coupling $g$ is
still rather small. During the past few years there have been a series
of conjectures about the behavior of string theory when the coupling
becomes much greater than one. It was previously believed
that there were five  fundamental
string theories in ten dimensions which differed in the amount of supersymmetry
and type of gauge groups they contained. The low energy limits of these
theories were ten dimensional supergravity coupled to matter.
In addition there was
an eleven dimensional supergravity theory which did not seem to fit into
string theory at all. It is now believed that all of these theories
are connected  in the sense that
the large coupling limit of one theory compactified on one type of internal
space is equivalent to the
weak coupling limit of another compactified on a possibly different
internal space. This suggests that there is one universal theory with different
weak coupling limits corresponding to each of the known theories.
The conjectures relating these theories are known as S-duality. The evidence
for them has been accumulating for the past two years 
and is now rather convincing \cite{pol2}.
This includes the fact that there are solitons in one theory with the
same properties as the fundamental strings of the other, 
and the spectrum
of BPS states in the two theories 
(which does not depend on the coupling) agrees.

I do not have time to discuss this exciting subject in detail,
but let me mention two consequences for our discussion of black holes.
For simplicity, I will not discuss compactification, but consider
black holes directly in ten dimensions.
Since all string theories include gravity, the Schwarzschild black hole
is a solution to each theory.  By taking different limits, one can
represent its states in different ways.
Let us consider the type IIB theory 
which has both fundamental strings and
D-strings. The D-strings have a tension $1/(g l_s^2)$ and hence are very heavy 
when $g \ll 1$. But when $g\gg 1$, the D-strings become  much lighter
than the fundamental strings. This limit is described by another
IIB string theory but now the role of the fundamental strings is played by
the D-strings and the coupling constant is $\hat g = 1/g$. (In other words,
the IIB theory is self-dual.) We saw in section 2 that when $g \ll 1$,
the states of a Schwarzschild black hole could be represented by states
of an ordinary string. We now see that
in the limit $g \gg 1$,  the same black hole can be described in terms of
states of a weakly coupled D-string. 

Thus one has the following picture as one increases $g$.
One can start with an excited state of the fundamental string at level $N$
when $g=0$.
As we have seen, when $g\sim 1/N^{1/4}$
the Schwarzschild radius is of order the string scale and
the state forms a black hole. If we continue to increase the coupling,
the black hole remains unchanged  until
$g \sim N^{1/4}$ which is when the Schwarzschild radius is of order the
length scale set by the D-string tension. Beyond this point,
the black hole can be described by an excited state of 
a weakly coupled D-string at the same level $N$ as the initial fundamental
string. 

In fact, the low energy IIB string lagrangian
has an SL(2,Z) symmetry, under which the Einstein metric is invariant,
but the fundamental string is mapped into a $(m,n)$ string which carries
$m$ units of fundamental string charge and $n$ units of D-string charge.
Starting with the Schwarzschild black hole, there are different weak
coupling limits of this theory in which the states are excitations of the
$(m,n)$ strings.

It is interesting to consider the black hole information puzzle 
 from this standpoint. If we compactify five dimensions of the IIB string theory
and wrap various D-branes around the internal five torus, one can obtain
extreme and near extreme five dimensional black holes.
We have seen that the spectrum of radiation at infinity
remains unchanged as we
increase the coupling and the description of the state changes from
slightly excited D-branes to near extremal black holes.
The D-brane radiation
is known to be unitary, yet it has been argued that the black hole
radiation will not be unitary. The duality conjectures imply that
if we continue to increase $g$,
the black hole can again we described by a weakly coupled dual string
theory. The spectrum will remain the same, and the radiation will again be
unitary. Of course the spacetime geometry in both limits $g\gg 1$ and
$g \ll 1$ is flat, so there are qualitative differences from the black hole.
But
still it seems rather unlikely that a physical process would change from
being unitary to nonunitary and back to unitary as a parameter is continuously
increased.

As a second application of duality ideas to black holes, we consider
the IIA string theory. This theory has a series of BPS states (bound
states of D-zerobranes) with
masses which are integer multiples of $1/(g l_s)$. This is similar to
the spectrum of Kaluza-Klein states in a theory compactified on a 
circle of radius $R = g l_s$. Note that for weak coupling, the radius is very
small, but at strong coupling, it becomes large. For this (and other) 
reasons, it is believed that the low energy limit of the strongly
coupled ten dimensional IIA string theory is eleven dimensional supergravity
compactified on a circle of radius $R = g l_s$. The eleven dimensional
Planck length turns out to be $l_p = g^{1/3} l_s$. One
can now trace out the following behavior of an excited IIA string state as
the coupling is increased. One starts at zero coupling with a
state at level $N$, so $M= \sqrt N/l_s$ and $S\sim \sqrt N$.
One now increases the string coupling keeping the state (i.e. the entropy)
fixed. As before,
when $g\sim 1/N^{1/4}$, one forms a ten dimensional black hole with
$r_0 = N^{1/16}$ in ten dimensional Planck units (since $S \sim r_0^8 \sim
N^{1/2}$). When $g\sim 1$, the length of the eleventh dimension becomes
greater than the eleven dimensional Planck length, and so becomes
physically meaningful. Since the ten dimensional IIA supergravity theory
is just the dimensional reduction of eleven dimensional supergravity,
the ten dimensional black hole then becomes
an eleven dimensional black string, i.e. the solution becomes the product
of Schwarzschild and a circle. As the coupling increases, the
length of the eleventh dimension becomes larger. 
When it becomes of
order $r_0$, the black string becomes unstable \cite{grla} and probably 
forms an eleven dimensional
black hole.  This occurs when $r_0 \sim R$, which implies $R
\sim S^{1/9} l_p$. Since $S\sim N^{1/2}$, we have $R \sim N^{1/18} l_p$
and hence $g\sim N^{1/12}$. As $g$ is increased further, corresponding 
to larger values of $R$, the state remains an eleven dimensional
black hole. Conversely, starting with an eleven dimensional Schwarzschild
solution with one dimension periodically identified, one can follow the
above description in reverse as one decreases $g$: The eleven
dimensional black hole transforms into an eleven dimensional black string,
which then becomes a ten dimensional black hole, and finally 
a weakly coupled string in ten dimensions. This is one approach toward
understanding the entropy of eleven dimensional black holes. Of course,
it would be more satisfactory to have a direct explanation of the
entropy of eleven dimensional
black holes, which did not require a  direction to be compactified.
But this requires the full eleven dimensional quantum theory (called M theory)
which is
not yet well understood.

\sect{Discussion}

Our understanding of black hole microstates provided by string theory is
progressing rapidly. To illustrate this, let me briefly list 
some of the highlights from 1996.\footnote{This is by no means an
exhaustive list of all of the contributions to this subject, which would
include well over a hundred papers.} In January, the first
calculation showing precise agreement between the entropy of an extremal
five dimensional black hole and the counting of string states was performed
\cite{stva}.
In February, this was extended to near extremal black holes \cite{cama,host2},
and extreme
rotating black holes \cite{bmpv} (still in five dimensions). In March, the
entropy of four dimensional black holes, both extremal \cite{mast1,jkm}
and near extremal \cite{hlm},
was reproduced. In June, the rate of low energy radiation from
a near extremal five dimensional black hole was shown  to agree exactly
with the rate from excited D-branes \cite{dama}. 
In August, this was extended to four 
dimensions \cite{gukl2}.
In September it was shown that the deviations from the black body
spectrum agree, both in five \cite{mast2} and four \cite{gukl} dimensions.
Finally, in December the entropy of black holes far from
extremality was understood, up to an overall coefficient, in terms of
a correspondence principle \cite{hopo}. 

Despite this enormous progress, our understanding is still far from 
complete. One outstanding issue is the resolution of the black hole
information puzzle, which was discussed in the last two sections. 
Perhaps a more modest question is
whether the entropy of black holes
far from extremality can be reproduced exactly. When the supersymmetric 
black hole calculations were first carried out, there was a strong
belief that supersymmetry was playing a key role, and it would
be impossible to do the same thing for nonsupersymmetric configurations.
But then it was found that the entropy continued to 
agree for near extremal black holes and also extremal black
holes which are not supersymmetric, e.g., extreme
rotating four dimensional black holes \cite{hlm,dab}. I now believe that the 
entropy of all large black holes should be computable, including the
overall coefficient.
The
correspondence principle certainly
shows that the main contribution to the entropy
can be understood without supersymmetry.

One might ask whether the correspondence principle could be extended to compare
the precise coefficients in the expression for the entropy. At first sight
this appears to be difficult since it requires a better understanding
of the string state when it is at the string scale, and interactions become
important. However,
there are indications that things are simpler than they appear.
One of these comes from studies of the near extremal threebrane.
Using the correspondence principle, one can understand the entropy of
all black $p$-branes in terms of a gas of massless strings on the brane
at weak coupling \cite{hopo}. However, for the threebrane, the entropy
turns out to be independent of when the masses are set equal. If one
takes the coupling all the way to zero and compares the entropy of a
free gas with that of the  black threebrane, one finds $S_{bh} = (3/4) S_{gas}$
\cite{gkp}.
The correspondence principle predicts that
the transition to the black $p$-brane occurs when the temperature
of the gas is of order the string scale, so the interactions should be
important. Since the potential energy is positive, 
this explains why $S_{bh} < S_{gas}$: 
By ignoring the interactions,
the energy of each state of the gas
has been underestimated, and hence the total number
of states with a given energy has been overestimated. But it is not clear
why the interactions  should simply produce a factor of $3/4$.

We have seen that the transition from a perturbative string state to
a black hole occurs when the string coupling is still rather small
$g\sim 1/ N^{1/4}$ or $g\sim 1/Q$ (for near extremal black holes).
So one might wonder why
string perturbation theory cannot be used to directly study properties of
black holes. The reason is that the effective coupling constant, due
to the long string or large number of D-branes, is really $g N^{1/4}$
or $gQ$, which is becoming of order one. Even though string perturbation
theory
is known to diverge \cite{grpe}, there is an important difference between
a coupling constant that is of order one, and one that is small.
The perturbation series has the form $\sum C_n \tilde g^n$ where
$C_n \sim (2n)!$ and $\tilde g$ is the effective coupling constant \cite{she}.
This is an asymptotic series, and the best approximation to the exact
answer is obtained by cutting the series off after $\tilde g^{-1/2}$ terms
when the individual terms become greater than one. 
It turns out that the error one introduces
this way is of order $e^{-1/\tilde g}$. (In ordinary field theory, $C_n \sim
n!$, so cutting the series off after $1/\tilde g$ terms introduces an error
$e^{-1/\tilde g^2}$.) Clearly, when $\tilde g \sim 1$, no useful information
can be obtained from the perturbation series. Fortunately, for $\tilde g \gg 1$,
one has an alternative description of the system in terms of a 
semi-classical black hole.

For the near extremal black holes discussed in section 3.2, the energy
above extremality is independent of the string coupling $g$. 
So one
can compare the entropy as a function of energy for the black holes and
strings. Since they agree, one knows that the temperature of the two
systems are equal. This is not true for black holes far from extremality.
We saw in section 2 that the mass of a state does not remain
constant as one changes the string coupling $g$. In string units, the
mass of a perturbative string state is approximately constant until it
forms a black hole and then it decreases with $g$. In
Planck units, the mass of a string state increases with $g$ until it
forms a black hole and then it remains constant. This means that even
though the entropies of black holes and strings agree, their temperatures
do not. For a highly excited string the natural temperature is a
constant of order the string scale  (the Hagedorn temperature) rather than
the Hawking temperature.

The understanding of black hole entropy provided by the correspondence
principle leads to a simple picture of the evaporation of a Schwarzschild
black hole, if the string coupling is small in nature. 
In most of our previous discussion, we imagined varying the
string coupling $g$. Now we suppose that $g \ll 1$ is fixed,
so the
string length scale is much larger than the Planck length.
A large black hole will Hawking evaporate until the curvature at the
horizon
reaches the string scale. At this point it turns into a highly
excited string state with $N \sim 1/g^4$, since for this value of $N$,
the string has a mass  comparable to the black hole: $M_s^2 \sim N/l_s^2 \sim
1/(g^4 l_s^2) \sim l_s^2/G^2 \sim M_{bh}^2$.  The entropy of the string is also
comparable to the black hole entropy at this point, so the string can
carry the remaining information in the black hole.
The excited string will then continue
to decay via string interactions. The temperature slowly increases as the black 
hole evaporates and reaches the string scale at the transition point. It then
remains at this temperature as the excited string decays.\footnote{The mass
is likely to decay exponentially if one assumes that each segment of the long
string has the same probability to break off a small loop of string. Since
$M\sim L$, $dM/dt \sim -M$. (I thank L. Susskind for pointing this out.)} 
Eventually
one is left with an unexcited string, i.e., an elementary particle
like a photon.

The history of particle physics is full of examples where objects which were
thought to be elementary were later found to be composed of more
fundamental entities.
The fact that the entropy of black holes has now been reproduced by
counting quantum states strongly suggests that we have finally identified
the fundamental degrees of freedom, and there is not another level
of structure waiting to be uncovered. However, as we saw in the  last
section, these fundamental degrees of freedom can take different forms in 
different weak coupling limits of the theory.

There remains the puzzling question of {\it why} the counting of string states
turns out to reproduce the area of the event horizon of a black hole.
This is undoubtably tied up with the fundamental question of
what is the origin of spacetime geometry in string theory. One recent 
suggestion, which is motivated by the relation between position in spacetime
and the moduli space of gauge theories on D-branes, 
starts with a quantum theory of $N\times N$ matrices in the limit of large $N$
\cite{bfss}. The description of black holes in this context is beginning
to be investigated \cite{lima,dvv}. At the risk of sounding too
conservative, I will state my belief that 
spacetime itself will ultimately be made up
of strings. 
After all, perturbations in the spacetime metric are just one
mode of the string. With the possible exception of zerobranes, it is difficult
to see how charged objects like D-branes can produce a neutral object like
Minkowski spacetime.
I believe the entropy calculations are providing a
glimpse into the quantum origin of spacetime. It is tempting to turn
the current calculations around and use them to try to {\it define} a metric or
area in spacetime in terms of the number of states in a Hilbert space.
Following this approach may expand our glimpse 
until
the full picture of quantum spacetime is finally revealed.

\sect{Acknowledgments}

It is a pleasure to thank A.~Ashtekar, T.~Jacobson, B.-L.~Hu, and S. Ross
for raising questions which (it is hoped) improved the clarity of this
presentation.  I also wish to thank J.~Polchinski, A.~Strominger, and 
L. Susskind for
discussions which improved my understanding of the results discussed
here.  This work was
supported in part by the NSF under grant PHY95-07065.


\begin{thebibliography}{99}
\baselineskip=16pt

\bibitem{chan} S. Chandrasekhar, {\it The Mathematical Theory of Black Holes},
Oxford University Press, Oxford (1983).
\bibitem{wald} R. Wald, gr-qc/9702022, in this volume.
\bibitem{anon} Anonymous
\bibitem{vafa} For a recent review  see C. Vafa, hep-th/9702201.
\bibitem{gsw} See e.g. M. Green, J. Schwarz and E. Witten, {\it
Superstring theory}, Cambridge University Press, Cambridge (1987).
\bibitem{mitu} D. Mitchell and N. Turok, \np B294, 1138, 1987.
\bibitem{suss1}
L. Susskind, hep-th/9309145.
\bibitem{hkrs} E. Halyo,  A. Rajaraman, and L. Susskind, \pl B392, 319, 1997,
hep-th/9605112;
E. Halyo, B. Kol, A. Rajaraman, and L. Susskind,  hep-th/9609075.
\bibitem{hopo} G. Horowitz and J. Polchinski, \pr D55, 6189, 1997,
hep-th/9612146.
\bibitem{kk} H. Leutwyler, {\sl Arch. Sci.} {\bf 13} (1960) 549;
P. Dobiasch and D. Maison, {\sl GRG} {\bf 14} (1982) 231;
A. Chodos and S. Detweiler, {\sl GRG} {\bf 14} (1982) 879;
D. Pollard, {\sl J.  Phys.} {\bf A 16} (1983) 565;
G. Gibbons and D. Wiltshire, {\sl Ann. Phys.} {\bf 167} (1986) 201, Erratum
\ap 176, 393, 1987.
\bibitem{lawi}
F. Larsen and F. Wilczek, {\sl Phys. Lett.} {\bf B375} (1996) 37,
hep-th/9511064; \np B475, 627, 1996,
hep-th/9604134; \np B488, 261, 1997, hep-th/9609084.
\bibitem{cvts} M. Cvetic and A. Tseytlin, {\sl Phys. Rev.}{\bf D53} (1996) 5619,
hep-th/9512031; A. Tseytlin, \np   B477, 431, 1996, hep-th/9605091.
\bibitem{dvv1} R. Dijkgraaf, E. Verlinde, and H. Verlinde, \np B486, 77, 1997,
hep-th/9603126; \np B486, 89, 1997, hep-th/9604055.

\bibitem{horo} G. Horowitz and S. Ross, hep-th/9704058.
\bibitem{host1} G. Horowitz and A. Strominger, \np B360, 197, 1991.
\bibitem{stva} A. Strominger and C. Vafa, \pl  B379, 99, 1996,
hep-th/9601029.
\bibitem{malthe} J. Maldacena, hep-th/9607235.
\bibitem{pol} J. Polchinski, \prl  75, 4724, 1995,
hep-th/9510017. For a recent
 review see J. Polchinski, {\it TASI Lectures on
 D-Branes,} ITP preprint NSF-ITP-96-145, hep-th/9611050.

\bibitem{dkps} M. Douglas, D. Kabat, P. Pouliot, and S. Shenker, 
\np B485, 85, 1997, hep-th/9608024.
\bibitem{con} A. Connes, {\it Noncommutative Geometry}, Academic Press,
(1994).
\bibitem{gihu} G. Gibbons and C. Hull, \pl 109B, 190, 1982.
\bibitem{n2bh} D. Kaplan, D. Lowe, J. Maldacena, and A. Strominger,
\pr D55, 4898, 1997, hep-th/9609204; K. Behrndt and T. Mohaupt,
hep-th/9611140; J. Maldacena, \pl B403, 20, 1997, hep-th/9611163;
W. Sabra, hep-th/9703101.

\bibitem{bmpv} J. Breckenridge, R. Myers, A. Peet and C. Vafa,
\pl B391, 93, 1997, hep-th/9602065.
\bibitem{mast1} J. Maldacena and A. Strominger, \prl 77, 428, 1996,
hep-th/9603060.
\bibitem{jkm} C. Johnson, R. Khuri, and R. Myers, \pl B378, 78, 1996,
hep-th/9603061.
\bibitem{homa} G. Horowitz and D. Marolf, \pr D55, 835, 1997, hep-th/9605244;
\pr 55, 846, 1997, hep-th/9606113.
\bibitem{hoya} G. Horowitz and H. Yang, hep-th/9701077.
\bibitem{hlm} G. Horowitz, D. Lowe, and  J. Maldacena, {\sl Phys. Rev. Lett.}
{\bf 77} (1996) 430, hep-th/9603195.
\bibitem{cvyo} M. Cvetic and D. Youm, hep-th/9508058; hep-th/9512127.
\bibitem{bala} V. Balasubramanian and F. Larsen, \np B478, 199, 1996,
hep-th/9604189.

\bibitem{cama} C. Callan and J. Maldacena, {\sl Nucl. Phys.}
{\bf B472} (1996) 591,
hep-th/9602043.
\bibitem{host2} G. Horowitz and A. Strominger, {\sl Phys. Rev. Lett.} {\bf 77} 
(1996) 2368, hep-th/9602051.
\bibitem{dama} S. Das and S. Mathur, {\sl Nucl. Phys.} {\bf B478} (1996) 561,
hep-th/9606185. 
\bibitem{mast2} J. Maldacena and A. Strominger, \pr D55, 861, 1997,
hep-th/9609026.
\bibitem{haw} S. Hawking, \cmp 43, 199, 1975.
\bibitem{stch} A. Starobinsky and S. Churilov,
{\sl Sov. Phys. JETP} {\bf 38} (1974) 1;
 G. Gibbons, {\sl Commun. Math. Phys.} {\bf 44} (1975) 245;
 D. Page, {\sl Phys. Rev.} {\bf D13} (1976) 198; {\sl Phys. Rev.}
{\bf D14} (1976) 3260;
 W. Unruh, {\sl Phys. Rev.} {\bf D14} (1976) 3251.

\bibitem{dgm} S. Das, G. Gibbons, and S. Mathur, \prl  78,
417, 1997, hep-th/9609052.
\bibitem{mast3} J. Maldacena and A. Strominger, hep-th/9702015.
\bibitem{mald} J. Maldacena, \pr D55, 7645, 1997, hep-th/9611125.
\bibitem{das} S. Das, hep-th/9703146.
\bibitem{gukl} S. Gubser and I. Klebanov, \prl 77, 4491, 1996, hep-th/9609076.
\bibitem{cgkt} C. Callan, S. Gubser, I. Klebanov, and A. Tseytlin,
\np B489, 65, 1997, hep-th/9610172.
\bibitem{krt} I. Klebanov, A. Rajaraman, and A. Tseytlin, hep-th/9704112.
\bibitem{deal} S. de Alwis and K. Sato, \pr D55, 6181, 1997, hep-th/9611189.

\bibitem{dkt} F. Dowker, D. Kastor and J. Traschen, hep-th/9702109.
\bibitem{hata} S. Hawking and M. Taylor-Robinson, hep-th/9702045.
\bibitem{krkl}  M. Krasnitz and I. Klebanov, hep-th/9703216.
\bibitem{page2} D. Page, \prl 71, 1291, 1993.
\bibitem{pol2} For a review, see
J. Polchinski, {\sl Rev. Mod. Phys.} {\bf 68} (1996) 1245,
hep-th/9607050.
\bibitem{grla} R. Gregory and R. Laflamme, \prl  70,
2837, 1993; \np B428, 399, 1994.

\bibitem{gukl2} S. Gubser and I. Klebanov, \np B482, 173, 1996, hep-th/9608108.
\bibitem{dab} A. Dabholkar, hep-th/9702050.
\bibitem{gkp} S. Gubser, I. Klebanov, and A. Peet, \pr D54, 3915, 1996,
hep-th/9602135; A. Strominger, unpublished.

\bibitem{grpe} D. Gross and V. Periwal, \prl 60, 2105, 1988.
\bibitem{she} S. Shenker, RU-90-047, 1990.
\bibitem{bfss} T. Banks, W. Fischler, S. Shenker, and L. Susskind,
\pr D55, 5112, 1997, hep-th/9610043.
\bibitem{lima} M. Li and E. Martinec, hep-th/9703211, hep-th/9704134.
\bibitem{dvv} R. Dijkgraaf, E. Verlinde, and H. Verlinde, hep-th/9704018.
\end{thebibliography}
\end{document}